\documentstyle[preprint,aps]{revtex}

\begin{document}
\title{ENERGY LEVELS AND WAVE FUNCTIONS OF VECTOR BOSONS IN HOMOGENEOUS MAGNETIC
FIELD}
\author{K.SOGUT, A.HAVARE, I.ACIKGOZ}
\address{MERSIN UNIVERSITY, DEPARTMENT OF PHYSICS}
\date{\today}
\maketitle

\begin{abstract}
We aimed to obtain the energy levels of spin-1 particles moving in a
constant magnetic field. The method used here is completely algebraic. In
the process to obtain the energy levels the wave function is choosen in
terms of Laguerre Polynomials.
\end{abstract}

\section{Introduction}

The problem of finding the states of a particle moving in an external field
has been solved for many situations. The relativistic wave equation of
spin-1 particles was at first derived by Kemmer in 1939 \cite{1}. This
equation is similar to the Dirac equation, but the technique investigated by
Kemmer is simpler. Kemmer equation was developed by different matrices.

There are various wave equations to describe the spin-1 particles given in
literature \cite{2,3,4,5}. The methods used in these equations are quite
complicated and give different results when they are compared. The method of
Shay and Good is to expand the six-component wave function in terms of
complete set of functions. This method leads to an invariant way to take
matrix elements, and it allows for arbitrary magnetic dipole moment and
electric quadrupole moment of the particle. The eigenvalues of Shay and
Good's equation have been obtained by Krase, Lu, Good by using the quite
complicated conventional method of solving the differential equation. A
general method of obtaining the energy levels of any spin theory has been
proposed by Tsai and Yildiz. They went on \ to obtain the energy levels of
spin-1 theory of Proca and Kemmer and observed that spin-1 theory is
consistent only when there is no anomalous -magnetic moment coupling.

The generally accepted method of approach is to define the Lagrange equation
of motion and use the differential equation technique to solve for the
eigenfunctions and eigenvalues of system. This method becomes complicated
when higher spinning particles are handled. In this study we used an
algebraic method for obtaining the energy levels of spin-1 particles. This
method of approach does not require an explicit solution of the equation.
For introducing the method we first used it to obtain the energy levels of
spin-$\frac{1}{2}$ particles in section two and by using the eigenfunctions
that found for spin-$\frac{1}{2}$ particles moving in terms of Laguerre
polynomials in Kemmer equation we obtained the energy levels of spin-1
particles in a homogeneous magnetic field. The vector potential we used is 
\begin{equation}
\overrightarrow{A}=\frac{1}{2}xB\widehat{y}-\frac{1}{2}yB\widehat{x}.
\end{equation}

\section{Energy Levels of Spin-$\frac{1}{2}$ Particles}

The wave equation is 
\begin{equation}
(\gamma ^{\mu }\pi _{\mu }-m)\Psi (x)=0\text{ }  \label{2}
\end{equation}
where $\gamma ^{\mu }$ are Dirac matrices and $\pi _{\mu }$ is
electodynamical four-momentum tensor. The explicit form of this equation is 
\begin{equation}
\left( 
\begin{array}{cccc}
\left( E-m\right)  & 0 & -P_{z} & -\pi _{-} \\ 
0 & \left( E-m\right)  & -\pi _{+} & P_{z} \\ 
-P_{z} & -\pi _{-} & \left( E-m\right)  & 0 \\ 
-\pi _{+} & P_{z} & 0 & \left( E-m\right) 
\end{array}
\right) \left( 
\begin{array}{c}
\varphi _{+} \\ 
\varphi _{-} \\ 
\chi _{+} \\ 
\chi _{-}
\end{array}
\right) =0  \label{3}
\end{equation}
where helicity raising and lowering operators are defined as 
\begin{eqnarray}
\pi _{+} &=&\pi _{1}+i\pi _{2}=e^{i\theta }\left[ -i\partial _{r}+\frac{1}{r}%
\partial _{\theta }-\frac{iBre}{2}\right] , \\
\pi _{-} &=&\pi _{1}-i\pi _{2}=e^{-i\theta }\left[ -i\partial _{r}-\frac{1}{r%
}\partial _{\theta }+\frac{iBre}{2}\right] .
\end{eqnarray}
Before obtaining the energy levels we must note that the effect of helicity
raising and lowering operators on Laguerre polynomials.

For m%
\mbox{$>$}%
0

\begin{eqnarray}
\pi _{+}\text{L}_{N}^{\left| m\right| }F_{m} &=&2i\omega ^{\frac{1}{2}}x^{%
\frac{1}{2}}\text{L}_{N-1}^{\left| m\right| +1}F_{m+1}, \\
\pi _{-}\text{L}_{N}^{\left| m\right| }F_{m} &=&-2iN\omega ^{\frac{1}{2}}x^{-%
\frac{1}{2}}\text{L}_{N+1}^{\left| m\right| -1}F_{m-1},
\end{eqnarray}
and for m%
\mbox{$<$}%
0

\begin{eqnarray}
\pi _{+}\text{L}_{N}^{\left| m\right| }F_{m} &=&-2i(N+\left| m\right|
)\omega ^{\frac{1}{2}}x^{-\frac{1}{2}}\text{L}_{N}^{\left| m\right|
-1}F_{m+1}, \\
\pi _{-}\text{L}_{N}^{\left| m\right| }F_{m} &=&2i\omega ^{\frac{1}{2}}x^{%
\frac{1}{2}}\text{L}_{N}^{\left| m\right| +1}F_{m-1},
\end{eqnarray}
where m is quantum number. Here \cite{7}; 
\begin{eqnarray}
N &=&\frac{1}{4\omega }\left( E^{2}-m^{2}-P_{z}^{2}\right) +\frac{\left(
m-\left| m\right| \right) }{2}\text{ ,\ \ \ \ \ \ \ for }m>0, \\
N &=&\frac{1}{4\omega }\left( E^{2}-m^{2}-P_{z}^{2}\right) +\frac{\left(
m-\left| m\right| -2\right) }{2}\text{ ,\ \ \ \ \ \ \ for }m<0
\end{eqnarray}
and 
\begin{eqnarray}
F_{m} &=&\left( \sqrt{\omega \rho ^{2}}\right) ^{\left| m\right|
}e^{im\theta +iP_{z}z-\frac{x}{2}}, \\
\omega  &=&\frac{eB}{2}\text{ \ \ , \ \ }x=\omega \rho ^{2}.
\end{eqnarray}
There are four situations due to spin orientations in magnetic field and
positive and negative values of m.

{\bf 1.Situation:} Spin is parallel to the magnetic field and m%
\mbox{$>$}%
0 \ \ \ \ \ \ \ \ 

If we choose the wave function in the form 
\begin{equation}
\Psi =\left[ 
\begin{array}{c}
\text{L}_{N}^{\left| m\right| }F_{m} \\ 
0 \\ 
c_{1}\text{L}_{N}^{\left| m\right| }F_{m} \\ 
d_{1}\text{L}_{N-1}^{\left| m\right| +1}F_{m+1}
\end{array}
\right] \text{ }
\end{equation}
\ and use it in Eq.(\ref{3}) we obtain energy spectrum and $c_{1},d_{1}$
coefficients as follows: 
\begin{eqnarray}
E_{\uparrow ,m>0}^{2} &=&P_{z}^{2}+m^{2}+4\omega \left( N-1\right) , \\
c_{1} &=&\frac{P_{z}}{\left( E+m\right) }\text{ \ ,} \\
d_{1} &=&\frac{2i\omega ^{\frac{1}{2}}x^{\frac{1}{2}}}{\left( E+m\right) }.
\end{eqnarray}

{\bf 2.Situation:} Spin is parallel to the magnetic field and m%
\mbox{$<$}%
0 \ \ \ \ \ \ \ \ 

If we choose the wave function in the form 
\begin{equation}
\Psi =\left[ 
\begin{array}{c}
\text{L}_{N}^{\left| m\right| }F_{m} \\ 
0 \\ 
c_{2}\text{L}_{N}^{\left| m\right| }F_{m} \\ 
d_{2}\text{L}_{N}^{\left| m\right| -1}F_{m+1}
\end{array}
\right] 
\end{equation}
and use it in Eq.(\ref{3}) we obtain energy spectrum $c_{2},$ $d_{2}$
coefficients as follows: 
\begin{eqnarray}
E_{\uparrow ,m<0}^{2} &=&P_{z}^{2}+m^{2}+4\omega \left( N+\left| m\right|
\right) , \\
c_{2} &=&\frac{P_{z}}{\left( E+m\right) }, \\
d_{2} &=&\frac{-2i\left( N+\left| m\right| \right) \omega ^{\frac{1}{2}}x^{-%
\frac{1}{2}}}{\left( E+m\right) }.
\end{eqnarray}

{\bf 3.Situation:} Spin is anti-parallel to the magnetic field and m%
\mbox{$>$}%
0 \ \ \ \ \ \ \ \ 

If we choose the wave function in the form 
\begin{equation}
\Psi =\left[ 
\begin{array}{c}
0 \\ 
\text{L}_{N}^{\left| m\right| }F_{m} \\ 
c_{3}\text{L}_{N+1}^{\left| m\right| -1}F_{m-1} \\ 
d_{3}\text{L}_{N}^{\left| m\right| }F_{m}
\end{array}
\right] 
\end{equation}
and use it in Eq.(\ref{3}) we obtain energy spectrum and $c_{3},$ $d_{3}$
coefficients as follows: 
\begin{eqnarray}
E_{\downarrow ,m>0}^{2} &=&P_{z}^{2}+m^{2}+4\omega N, \\
c_{3} &=&\frac{-2iN\omega ^{\frac{1}{2}}x^{-\frac{1}{2}}}{\left( E+m\right) }%
, \\
d_{3} &=&\frac{-P_{z}}{\left( E+m\right) }.
\end{eqnarray}
{\bf 4.Situation:} Spin is anti-parallel to the magnetic field and m%
\mbox{$<$}%
0 \ \ \ \ \ \ \ \ 

If we choose the wave function in the form 
\begin{equation}
\Psi =\left[ 
\begin{array}{c}
0 \\ 
\text{L}_{N}^{\left| m\right| }F_{m} \\ 
c_{4}\text{L}_{N}^{m+1}F_{m-1} \\ 
d_{4}\text{L}_{N}^{\left| m\right| }F_{m}
\end{array}
\right] 
\end{equation}
and use it in Eq.(\ref{3}) we obtain energy spectrum and $c_{4},$ $d_{4}$
coefficients as follows: 
\begin{eqnarray}
E_{\downarrow ,m<0}^{2} &=&P_{z}^{2}+m^{2}+4\omega (N+\left| m\right| +1, \\
c_{4} &=&\frac{2i\omega ^{\frac{1}{2}}x^{\frac{1}{2}}}{\left( E+m\right) },
\\
d_{4} &=&\frac{-P_{z}}{\left( E+m\right) }.
\end{eqnarray}

\section{Eigenvalue Equation and Energy Spectrum of Spin-1 Particle}

Massive spin-1 particle will be considered as a two-particle system of spin-$%
\frac{1}{2}$ instead of a single spin-1 particle. This is the second
quantization approach of Schr\"{o}dinger to the problem. For a massive
spin-1 particle the Kemmer equation is given in the form 
\begin{equation}
\left( \beta ^{\mu }\pi _{\mu }-m\right) \Psi \left( x\right) =0
\end{equation}
where 16$\times $16 Kemmer matrices $\beta ^{\mu }$ are given as 
\[
\beta ^{\mu }=\gamma ^{\mu }\otimes \text{I +I}\otimes \gamma ^{\mu } 
\]
with usual $\gamma ^{\mu }$ Dirac matrices. $\pi _{\mu }$ is
electrodynamical four momentum tensor and m is mass of two spin-$\frac{1}{2}$
particles. The geometry of the problem is cylindrically since the particle
is rotating around z-axis with a $\theta $ angle in xy-plane and going
forward to z-axis direction.

The sixteen-component wave function of the equation can be written in the
form 
\[
\Psi =\left( 
\begin{array}{c}
A \\ 
B \\ 
C \\ 
D
\end{array}
\right) 
\]
where A, B, C, D are 4-component spinors and given as follows: 
\begin{eqnarray*}
A &=&\left( 
\begin{array}{c}
A_{+} \\ 
A_{0} \\ 
A_{\tilde{o}} \\ 
A_{-}
\end{array}
\right) =\left( 
\begin{array}{c}
\varphi _{+}\varphi _{+} \\ 
\varphi _{+}\varphi _{-} \\ 
\varphi _{-}\varphi _{+} \\ 
\varphi _{-}\varphi _{-}
\end{array}
\right) \text{ \ \ \ \ },B=\left( 
\begin{array}{c}
B_{+} \\ 
B_{0} \\ 
B_{\tilde{o}} \\ 
B_{-}
\end{array}
\right) =\left( 
\begin{array}{c}
\varphi _{+}\chi _{+} \\ 
\varphi _{+}\chi _{-} \\ 
\varphi _{-}\chi _{+} \\ 
\varphi _{-}\chi _{-}
\end{array}
\right)  \\
C &=&\left( 
\begin{array}{c}
C_{+} \\ 
C_{0} \\ 
C_{\tilde{o}} \\ 
C_{-}
\end{array}
\right) =\left( 
\begin{array}{c}
\chi _{+}\varphi _{+} \\ 
\chi _{+}\varphi _{-} \\ 
\chi _{-}\varphi _{+} \\ 
\chi _{-}\varphi _{-}
\end{array}
\right) \text{ \ \ \ \ },D=\left( 
\begin{array}{c}
D_{+} \\ 
D_{0} \\ 
D_{\tilde{o}} \\ 
D_{-}
\end{array}
\right) =\left( 
\begin{array}{c}
\chi _{+}\chi _{+} \\ 
\chi _{+}\chi _{-} \\ 
\chi _{-}\chi _{+} \\ 
\chi _{-}\chi _{-}
\end{array}
\right) .
\end{eqnarray*}
In these matrices the left and right elements of the right matrix indicate
first and second particles respectively and ``+'' and ``-'' indices indicate
the spin orientations in magnetic field.

After some algebra Kemmer equation takes the following form 
\begin{equation}
\left[ \left( \gamma _{1}^{0}\otimes \text{I}_{2}+\text{I}_{1}\otimes \gamma
_{2}^{0}\right) E+\left( \gamma _{1}^{0}\otimes \overrightarrow{\alpha }_{2}+%
\overrightarrow{\alpha }_{1}\otimes \gamma _{2}^{0}\right) \cdot 
\overrightarrow{\pi }-m\gamma _{1}^{0}\otimes \gamma _{2}^{0}\right] \left( 
\begin{array}{c}
A \\ 
B \\ 
C \\ 
D
\end{array}
\right) =0  \label{31}
\end{equation}
where 
\begin{eqnarray*}
\overrightarrow{\alpha } &=&\gamma ^{0}\overrightarrow{\gamma }=\left( 
\begin{array}{cc}
0 & \overrightarrow{\sigma } \\ 
\overrightarrow{\sigma } & 0
\end{array}
\right) , \\
\gamma ^{0} &=&\left( 
\begin{array}{cc}
\text{I} & 0 \\ 
0 & -\text{I}
\end{array}
\right) .
\end{eqnarray*}
We obtain four linear algebraic equations, ultimately from Eq.(\ref{31}): 
\begin{eqnarray}
\left( 2E-m\right) A-\overrightarrow{\sigma }_{(1)}\cdot \overrightarrow{\pi 
}C-\overrightarrow{\sigma }_{_{\left( 2\right) }}\cdot \overrightarrow{\pi }%
B &=&0,  \label{32} \\
\overrightarrow{\sigma }_{(2)}\cdot \overrightarrow{\pi }A-\overrightarrow{%
\sigma }_{(1)}\cdot \overrightarrow{\pi }D-mB &=&0,  \label{33} \\
\overrightarrow{\sigma }_{(1)}\cdot \overrightarrow{\pi }A-\overrightarrow{%
\sigma }_{(2)}\cdot \overrightarrow{\pi }D-mC &=&0,  \label{34} \\
\left( 2E+m\right) D-\overrightarrow{\sigma }_{(1)}\cdot \overrightarrow{\pi 
}B-\overrightarrow{\sigma }_{(2)}\cdot \overrightarrow{\pi }C &=&0,
\label{35}
\end{eqnarray}
where 
\begin{eqnarray}
\overrightarrow{\sigma }_{\left( 1\right) }\cdot \overrightarrow{\pi }
&=&\left( \overrightarrow{\sigma }\otimes \text{I}\right) \cdot 
\overrightarrow{\pi }=\left( 
\begin{array}{cccc}
P_{z} & 0 & \pi _{-} & 0 \\ 
0 & P_{z} & 0 & \pi _{-} \\ 
\pi _{+} & 0 & -P_{z} & 0 \\ 
0 & \pi _{+} & 0 & -P_{z}
\end{array}
\right) ,  \label{36} \\
\overrightarrow{\sigma }_{\left( 2\right) }\cdot \overrightarrow{\pi }
&=&\left( \text{I}\otimes \overrightarrow{\sigma }\right) \cdot 
\overrightarrow{\pi }=\left( 
\begin{array}{cccc}
P_{z} & \pi _{-} & 0 & 0 \\ 
\pi _{+} & -P_{z} & 0 & 0 \\ 
0 & 0 & P_{z} & \pi _{-} \\ 
0 & \pi _{+} & \pi _{+} & -P_{z}
\end{array}
\right) ,  \label{37}
\end{eqnarray}
and 
\begin{eqnarray}
\pi _{+} &=&\pi _{1}+i\pi _{2}=e^{i\theta }\left[ -i\partial _{r}+\frac{1}{r}%
\partial _{\theta }-\frac{iBre}{2}\right] , \\
\pi _{-} &=&\pi _{1}-i\pi _{2}=e^{-i\theta }\left[ -i\partial _{r}-\frac{1}{r%
}\partial _{\theta }+\frac{iBre}{2}\right] .
\end{eqnarray}
By using the explicit forms\ of matrices seen in Eqs.(\ref{36}), (\ref{37})
and substituting the following equalities 
\begin{eqnarray*}
B_{0}\left( \varphi _{+}\chi _{-}\right)  &=&C_{\tilde{o}}\left( \chi
_{-}\varphi _{+}\right) \text{ \ \ \ \ \ ,}B_{\tilde{o}}\left( \varphi
_{-}\chi _{+}\right) =C_{0}\left( \chi _{+}\varphi _{-}\right) , \\
A_{0}\left( \varphi _{+}\varphi _{-}\right)  &=&A_{\tilde{o}}\left( \varphi
_{-}\varphi _{+}\right) \text{ \ \ \ \ \ ,}D_{0}\left( \chi _{+}\chi
_{-}\right) =D_{\tilde{o}}\left( \chi _{-}\chi _{+}\right) ,
\end{eqnarray*}
into four equations given by Eqs.(\ref{32})-(\ref{35}) we obtain ten linear
algebraic equations in the form 
\begin{eqnarray}
\left( 2E-m\right) A_{+}-2P_{z}B_{+}-2\pi _{-}B_{0} &=&0,  \label{40} \\
\left( 2E-m\right) A_{0}-\pi _{+}B_{+}-\pi _{-}B_{-}+P_{z}\left( B_{0}-B_{%
\tilde{o}}\right)  &=&0,  \label{41} \\
\left( 2E-m\right) A_{-}+2P_{z}B_{-}-2\pi _{+}B_{\tilde{o}} &=&0,  \label{42}
\\
\left( 2E-m\right) D_{+}-2P_{z}B_{+}-2\pi _{-}B_{\tilde{o}} &=&0,  \label{43}
\\
\left( 2E-m\right) D_{0}-\pi _{+}B_{+}-\pi _{-}B_{-}+P_{z}\left( B_{\tilde{o}%
}-B_{0}\right)  &=&0,  \label{44} \\
\left( 2E-m\right) D_{-}+2P_{z}B_{-}-2\pi _{+}B_{0} &=&0,  \label{45} \\
P_{z}\left( A_{+}-D_{+}\right) +\pi _{-}\left( A_{0}-D_{0}\right) -mB_{+}
&=&0,  \label{46} \\
P_{z}\left( D_{-}-A_{-}\right) +\pi _{+}\left( A_{0}-D_{0}\right) -mB_{-}
&=&0,  \label{47} \\
P_{z}\left( A_{0}+D_{0}\right) +\pi _{-}A_{-}-\pi _{+}D_{+}-mB_{\tilde{o}}
&=&0,  \label{48} \\
P_{z}\left( A_{0}+D_{0}\right) -\pi _{+}A_{+}+\pi _{-}D_{-}+mB_{0} &=&0.
\label{49}
\end{eqnarray}
In order to find the energy eigenvalues we will choose B$_{0}$ and B$_{%
\tilde{o}}$ in terms of Laguerre polynomials: 
\begin{eqnarray*}
B_{0} &=&\text{L}_{N}^{\left| m\right| }F_{m}, \\
B_{\tilde{o}} &=&-\text{L}_{N}^{\left| m\right| }F_{m}.\text{ \ }
\end{eqnarray*}

$\left( i\right) $ for m%
\mbox{$>$}%
0

If we use $B_{0}=$L$_{N}^{\left| m\right| }F_{m}$ and $B_{\tilde{o}}=-$L$%
_{N}^{\left| m\right| }F_{m}$ \ in Eqs.(\ref{40})-(\ref{49}) we find A, B,
C, D spinors in the form
\begin{eqnarray*}
B &=&\left( 
\begin{array}{c}
b_{+}\omega ^{\frac{1}{2}}x^{-\frac{1}{2}}L_{N+1}^{\left| m\right| -1}F_{m-1}
\\ 
L_{N}^{\left| m\right| }F_{m} \\ 
-L_{N}^{\left| m\right| }F_{m} \\ 
b_{-}\omega ^{\frac{1}{2}}x^{\frac{1}{2}}L_{N-1}^{\left| m\right| +1}F_{m+1}
\end{array}
\right) \text{ \ \ \ \ \ \ \ \ \ },C=\left( 
\begin{array}{c}
b_{+}\omega ^{\frac{1}{2}}x^{-\frac{1}{2}}L_{N+1}^{\left| m\right| -1}F_{m-1}
\\ 
-L_{N}^{\left| m\right| }F_{m} \\ 
L_{N}^{\left| m\right| }F_{m} \\ 
b_{-}\omega ^{\frac{1}{2}}x^{\frac{1}{2}}L_{N-1}^{\left| m\right| +1}F_{m+1}
\end{array}
\right) , \\
A &=&\left( 
\begin{array}{c}
\frac{2}{2E-m}\left( P_{z}b_{+}-2iN\right) \omega ^{\frac{1}{2}}x^{-\frac{1}{%
2}}L_{N+1}^{\left| m\right| -1}F_{m-1} \\ 
\frac{1}{2E-m}\left[ 2i\omega \left( b_{+}-\left( N-1\right) b_{-}\right)
-2P_{z}\right] L_{N}^{\left| m\right| }F_{m} \\ 
\frac{1}{2E-m}\left[ 2i\omega \left( b_{+}-\left( N-1\right) b_{-}\right)
-2P_{z}\right] L_{N}^{\left| m\right| }F_{m} \\ 
-\frac{2}{2E-m}\left( P_{z}b_{-}+2i\right) \omega ^{\frac{1}{2}}x^{\frac{1}{2%
}}L_{N-1}^{\left| m\right| +1}F_{m+1}
\end{array}
\right) , \\
D &=&\left( 
\begin{array}{c}
\frac{2}{2E+m}\left( P_{z}b_{+}+2iN\right) \omega ^{\frac{1}{2}}x^{-\frac{1}{%
2}}L_{N+1}^{\left| m\right| -1}F_{m-1} \\ 
\frac{1}{2E+m}\left[ 2i\omega \left( b_{+}-\left( N-1\right) b_{-}\right)
+2P_{z}\right] L_{N}^{\left| m\right| }F_{m} \\ 
\frac{1}{2E+m}\left[ 2i\omega \left( b_{+}-\left( N-1\right) b_{-}\right)
+2P_{z}\right] L_{N}^{\left| m\right| }F_{m} \\ 
-\frac{2}{2E+m}\left( P_{z}b_{-}-2i\right) \omega ^{\frac{1}{2}}x^{\frac{1}{2%
}}L_{N-1}^{\left| m\right| +1}F_{m+1}
\end{array}
\right) .
\end{eqnarray*}
We obtain four algebraic equations by using these spinors in Eqs.(\ref{46})-(%
\ref{49}) in the form 
\begin{eqnarray}
\left[ E^{2}-P_{z}^{2}-\frac{m^{2}}{4}-2\omega N\right] b_{+}+2N\left(
N-1\right) \omega b_{-} &=&0,\text{\ \ \ \ \ \ \ \ }  \label{50} \\
\left[ E^{2}-P_{z}^{2}-\frac{m^{2}}{4}-2\omega (N-1)\right] b_{-}+2\omega
b_{+} &=&0,\text{ \ \ \ \ \ \ \ \ \ \ }  \label{51} \\
\left[ E^{2}-P_{z}^{2}-\frac{m^{2}}{4}+4\omega \left( \frac{E}{m}-\left( N-%
\frac{1}{2}\right) \right) \right] +i\omega P_{z}\left( N-1\right)
b_{-}+i\omega P_{z}b_{+} &=&0,\text{ \ \ \ \ \ \ \ \ \ }  \label{52} \\
\left[ E^{2}-P_{z}^{2}-\frac{m^{2}}{4}-4\omega \left( \frac{E}{m}+\left( N-%
\frac{1}{2}\right) \right) \right] -i\omega P_{z}\left( N-1\right)
b_{-}-i\omega P_{z}b_{+} &=&0.\text{ \ \ \ \ \ \ \ \ \ }  \label{53}
\end{eqnarray}
From Eqs.(\ref{50})-(\ref{52}) we find 
\begin{eqnarray}
b_{+} &=&\frac{2N\omega \left[ E^{2}-P_{z}^{2}-\frac{m^{2}}{4}+4\omega
\left( \frac{E}{m}-\left( N-\frac{1}{2}\right) \right) \right] }{i\omega
P_{z}\left[ E^{2}-P_{z}^{2}-\frac{m^{2}}{4}-4\omega N\right] }, \\
b_{-} &=&\frac{-\left[ E^{2}-P_{z}^{2}-\frac{m^{2}}{4}+4\omega \left( \frac{E%
}{m}-\left( N-\frac{1}{2}\right) \right) \right] \left[ E^{2}-P_{z}^{2}-%
\frac{m^{2}}{4}-2\omega N\right] }{i\omega P_{z}\left( N-1\right) \left[
E^{2}-P_{z}^{2}-\frac{m^{2}}{4}-4\omega N\right] .}.
\end{eqnarray}
If we substitute these into Eqs.(\ref{51})-(\ref{53}) we obtain energy
levels as follows 
\begin{eqnarray}
E^{2} &=&P_{z}^{2}+\frac{m^{2}}{4}+4\omega \left( N-\frac{1}{2}\right) , \\
E &=&\frac{2\omega }{m}\pm \frac{1}{2}\sqrt{\frac{16\omega ^{2}}{m^{2}}%
+m^{2}+4P_{z}^{2}+16\omega \left( N-\frac{1}{2}\right) }, \\
E &=&-\frac{2\omega }{m}\pm \frac{1}{2}\sqrt{\frac{16\omega ^{2}}{m^{2}}%
+m^{2}+4P_{z}^{2}+16\omega \left( N-\frac{1}{2}\right) .}
\end{eqnarray}

$\left( ii\right) $ for m%
\mbox{$<$}%
0

If we use $B_{0}=$L$_{N}^{\left| m\right| }F_{m}$ and $B_{\tilde{o}}=-$L$%
_{N}^{\left| m\right| }F_{m}$ \ in Eqs.(\ref{40})-(\ref{49}) we find A, B,
C, D spinors in the form 
\begin{eqnarray*}
B &=&\left( 
\begin{array}{c}
b_{+}\omega ^{\frac{1}{2}}x^{\frac{1}{2}}L_{N}^{\left| m\right| +1}F_{m-1}
\\ 
L_{N}^{\left| m\right| }F_{m} \\ 
-L_{N}^{\left| m\right| }F_{m} \\ 
b_{-}\omega ^{\frac{1}{2}}x^{-\frac{1}{2}}L_{N}^{\left| m\right| -1}F_{m+1}
\end{array}
\right) \text{ \ \ \ \ \ \ \ \ \ },C=\left( 
\begin{array}{c}
b_{+}\omega ^{\frac{1}{2}}x^{\frac{1}{2}}L_{N}^{\left| m\right| +1}F_{m-1}
\\ 
-L_{N}^{\left| m\right| }F_{m} \\ 
L_{N}^{\left| m\right| }F_{m} \\ 
b_{-}\omega ^{\frac{1}{2}}x^{-\frac{1}{2}}L_{N}^{\left| m\right| -1}F_{m+1}
\end{array}
\right) , \\
A &=&\left( 
\begin{array}{c}
\frac{2}{2E-m}\left( P_{z}b_{+}+2i\right) \omega ^{\frac{1}{2}}x^{\frac{1}{2}%
}L_{N}^{\left| m\right| +1}F_{m-1} \\ 
\frac{1}{2E-m}\left[ -2i\omega \left( N+\left| m\right| +1\right)
b_{+}+2i\omega b_{-}-2P_{z}\right] L_{N}^{\left| m\right| }F_{m} \\ 
\frac{1}{2E-m}\left[ -2i\omega \left( N+\left| m\right| +1\right)
b_{+}+2i\omega b_{-}-2P_{z}\right] L_{N}^{\left| m\right| }F_{m} \\ 
\frac{2}{2E-m}\left[ -P_{z}b_{-}+2i\left( N+\left| m\right| \right) \right]
\omega ^{\frac{1}{2}}x^{-\frac{1}{2}}L_{N}^{\left| m\right| -1}F_{m+1}
\end{array}
\right) , \\
D &=&\left( 
\begin{array}{c}
\frac{2}{2E-m}\left( P_{z}b_{+}-2i\right) \omega ^{\frac{1}{2}}x^{\frac{1}{2}%
}L_{N}^{\left| m\right| +1}F_{m-1} \\ 
\frac{1}{2E-m}\left[ -2i\omega \left( N+\left| m\right| +1\right)
b_{+}+2i\omega b_{-}+2P_{z}\right] L_{N}^{\left| m\right| }F_{m} \\ 
\frac{1}{2E-m}\left[ -2i\omega \left( N+\left| m\right| +1\right)
b_{+}+2i\omega b_{-}+2P_{z}\right] L_{N}^{\left| m\right| }F_{m} \\ 
-\frac{2}{2E-m}\left[ P_{z}b_{-}+2i\left( N+\left| m\right| \right) \right]
\omega ^{\frac{1}{2}}x^{-\frac{1}{2}}L_{N}^{\left| m\right| -1}F_{m+1}
\end{array}
\right) .
\end{eqnarray*}
We obtain four algebraic equations by using these spinors in Eqs.(\ref{50})-(%
\ref{54}) in the form
\begin{equation}
\left[ E^{2}-P_{z}^{2}-\frac{m^{2}}{4}-2\omega \left( N+\left| m\right|
+1\right) \right] b_{+}+2\omega b_{-}=0
\end{equation}
\begin{equation}
\left[ E^{2}-P_{z}^{2}-\frac{m^{2}}{4}+4\omega \left( \frac{E}{m}-(N+\left|
m\right| +\frac{1}{2})\right) \right] -i\omega P_{z}b_{-}-i\omega \left(
N+\left| m\right| +\frac{1}{2}\right) P_{z}b_{+}=0
\end{equation}
\begin{equation}
\left[ E^{2}-P_{z}^{2}-\frac{m^{2}}{4}-4\omega \left( \frac{E}{m}+(N+\left|
m\right| +\frac{1}{2})\right) \right] +i\omega P_{z}b_{-}+i\omega \left(
N+\left| m\right| +\frac{1}{2}\right) P_{z}b_{+}=0
\end{equation}
\begin{equation}
\left[ E^{2}-P_{z}^{2}-\frac{m^{2}}{4}-2\omega (N+\left| m\right| )\right]
b_{-}+2\omega \left( N+\left| m\right| \right) \left( N+\left| m\right|
+1\right) b_{+}=0
\end{equation}
From Eqs.(\ref{50})-(\ref{52}) we find 
\begin{eqnarray}
b_{+} &=&\frac{2\omega \left[ E^{2}-P_{z}^{2}-\frac{m^{2}}{4}-4\omega \left( 
\frac{E}{m}+\left( N+\left| m\right| +\frac{1}{2}\right) \right) \right] }{%
i\omega P_{z}\left[ E^{2}-P_{z}^{2}-\frac{m^{2}}{4}-4\omega \left( N+\left|
m\right| +1\right) \right] }, \\
b_{-} &=&\frac{-\left[ E^{2}-P_{z}^{2}-\frac{m^{2}}{4}-4\omega \left( \frac{E%
}{m}+\left( N+\left| m\right| +\frac{1}{2}\right) \right) \right] \left[
E^{2}-P_{z}^{2}-\frac{m^{2}}{4}-2\omega \left( N+\left| m\right| +1\right) 
\right] }{i\omega P_{z}\left[ E^{2}-P_{z}^{2}-\frac{m^{2}}{4}-4\omega \left(
N+\left| m\right| +1\right) \right] },
\end{eqnarray}
If we substitute these into Eqs.(\ref{51})-(\ref{53}) we obtain energy
levels as follows 
\begin{eqnarray}
E^{2} &=&P_{z}^{2}+\frac{m^{2}}{4}+4\omega \left( N+\left| m\right| +\frac{1%
}{2}\right) , \\
E &=&\frac{2\omega }{m}\pm \frac{1}{2}\sqrt{\frac{16\omega ^{2}}{m^{2}}%
+m^{2}+4P_{z}^{2}+16\omega \left( N+\left| m\right| +\frac{1}{2}\right) ,} \\
E &=&-\frac{2\omega }{m}\pm \frac{1}{2}\sqrt{\frac{16\omega ^{2}}{m^{2}}%
+m^{2}+4P_{z}^{2}+16\omega \left( N+\left| m\right| +\frac{1}{2}\right) .}
\end{eqnarray}

\section{Conclusions}

There are a few techniques developed for understanding the behavior of
spin-1 particles moving in a constant magnetic field. These techniques are
quite complicated. Because of no available experimental tests of these
techniques we do not know which is the best.

We used here a simple algebraic method for obtaining the energy levels of
spin-1 particle moving in a constant magnetic field. Spin-1 particle was
considered as a two-fermion system and in the process to obtain the energy
levels the wave functions of spin-$\frac{1}{2}$ particle which are written
in terms of Laguerre polynomials were used.

If the magnetic field is quite weak, we see that the difference of the
second and third from the first energy level is $\pm \frac{2\omega }{m}.$
This comes from the spin orientations in magnetic field. The second and
third energy spectrums can be interpreted as the energy spectrums of
situations in which the spins of particle and anti-particle are parallel and
anti-parallel to the magnetic field, respectively. From these spectrums it
is seen for all values of magnetic field the transition energy of spin from
s=1 to s=-1 is equal to 
\[
\Delta E_{\pm }=\frac{4\omega }{m}\text{ \ \ .}
\]

\qquad

\qquad \qquad

\end{document}